\documentclass[twocolumn,superscriptaddress,aps,prb,longbibliography]{revtex4-2}
\usepackage[normalem]{ulem}
\usepackage{graphicx}
\usepackage{bm}
\usepackage{epstopdf}
\usepackage{amsmath}
\usepackage{amssymb}
\usepackage[ansinew]{inputenc}
\usepackage{multirow}
\usepackage[urlcolor=blue,colorlinks=true,citecolor=blue,linkcolor=blue,
            bookmarks=false]{hyperref}

\graphicspath{{Figures/}{../Figures/}{.}}
\newcommand{\seb}{Sm$_{1-x}$Eu$_x$B$_6$}

\usepackage{xcolor}
\usepackage[OT4]{fontenc}

\begin{document}
\title{Magnetic and electronic inhomogeneity in Sm$_{1-x}$Eu$_x$B$_6$}

\author{M. Victoria Ale Crivillero}
\affiliation{Max Planck Institute for Chemical Physics of Solids, D-01187
Dresden, Germany}

\author{Priscila F. S. Rosa}
\affiliation{Los Alamos National Laboratory, Los Alamos, NM 87545, USA}

\author{Z. Fisk}
\affiliation{Department of Physics and Astronomy, UC Irvine, Irvine, CA
92697, USA}

\author{J. M\"{u}ller}
\affiliation{Institute of Physics, Goethe-University Frankfurt, 60438
Frankfurt (M), Germany}

\author{P. Schlottmann}
\affiliation{Department of Physics, Florida State University, Tallahassee,
Florida 32306, USA}

\author{S. Wirth}
\email[e-mail: ]{steffen.wirth@cpfs.mpg.de}
\affiliation{Max Planck Institute for Chemical Physics of Solids, D-01187
Dresden, Germany}

\date{\today}

\begin{abstract}
While SmB$_6$ attracts attention as a possible topological Kondo insulator,
EuB$_6$ is known to host magnetic polarons that give rise to large
magnetoresistive effects above its ferromagnetic order transition. Here we
investigate single crystals of Sm$_{1-x}$Eu$_x$B$_6$ by magnetic and
magnetotransport measurements to explore a possible interplay of these two
intriguing phenomena, with focus on the Eu-rich substitutions.
Sm$_{0.01}$Eu$_{0.99}$B$_6$ exhibits generally similar behavior as EuB$_6$.
Interestingly, Sm$_{0.05}$Eu$_{0.95}$B$_6$ combines global antiferromagnetic
order with local polaron formation. A pronounced hysteresis is found in the
magnetoresistance of Sm$_{0.1}$Eu$_{0.9}$B$_6$ at low temperature ($T=$ 1.9~K)
and applied magnetic fields between 2.3 -- 3.6~T. The latter is in agreement
with a phenomenological model that predicts the stabilization of ferromagnetic
polarons with increasing magnetic field within materials with global
antiferromagnetic order.
\end{abstract}
\maketitle

\section{Introduction}
Hexaborides $R$B$_6$ of the cubic structure type CaB$_6$ \cite{pau34} form for
quite a number of elements $R$ \cite{eto77,ino21b}, primarily including
alkaline earths or rare earths (RE), and Y. They typically feature high melting
points, high hardness and excellent chemical stability \cite{cah19}. In
addition, this class of compounds is known for a remarkable variety of
properties ranging from the low work function of LaB$_6$ \cite{ges84} to
antiferromagnetic behavior in PrB$_6$, NdB$_6$ and GdB$_6$ \cite{geb68},
quadrupolar order in CeB$_6$ \cite{eff85} and even superconductivity below
7.1~K in YB$_6$ \cite{sch82}. In general, the B-octahedra take up two electrons
such that hexaborides formed with trivalent elements are typically highly
conductive while those based on divalent elements exhibit very low charge
carrier densities \cite{gru85}. An interesting exception here is SmB$_6$ with
its intermediate and temperature dependent valence \cite{vai64,lut16,zab18} of
around 2.6 at room temperature.

SmB$_6$ attracted enormous interest as a correlated topological insulator
candidate \cite{dze10,li20,wir21}. The existence of surface states on SmB$_6$
is well established \cite{kim13,mat20}. Their origin due to non-trivial
topology remains---despite overwhelming evidence (see e.g.\ \cite{li20} and
references therein)---controversially discussed \cite{hla18}. Nonetheless, the
surface states, once formed on the highly insulating bulk at low enough
temperatures \cite{eo19}, can be considered an electronic inhomogeneity (with
respect to the bulk) with potential applications. Indeed, topological
insulators with their spin-polarized surface states are considered as
candidates for next-generation spintronics \cite{wan16}.

Another type of electronic inhomogeneity, also considered useful for
applications, are magnetic polarons \cite{coe99,mol01,mol07}. Here, a magnetic
polaron corresponds to an entity made up by a charge carrier and local magnetic
moments within which strong exchange coupling localizes the charge carrier as
well as it aligns the magnetic moments locally in a ferromagnetic fashion.
Eu$^{2+}$ with its half-filled 4$f$ shell often provides the required exchange
interaction \cite{kun05}. In addition, divalent Eu$^{2+}$ in EuB$_6$
establishes a very low carrier concentration (as mentioned above) such that
separated polarons instead of extended bands may form above the global
ferromagnetic ordering transition at around 12~K \cite{nyh97,sue98,zha09,poh18,
bea22a}. In general, early indications for developing the concept of
polaron formation \cite{kas68} were provided by magnetotransport measurements
\cite{mol67,mol68} which are still heavily used in this context. Further
evidence for magnetic polarons came from more elaborate techniques including
muon spin-relaxation \cite{sto09}, neutron spectroscopy \cite{ter97} and
scanning tunneling microscopy (STM) \cite{ron06,poh18}. The latter gave an
approximate extension of the polarons of order few nanometers in layered
manganites and EuB$_6$, respectively.

The crystal structure of intertwined cubes of $R$ and B-octahedra in the
hexaborides $R$B$_6$ allows not only numerous elements to be placed on the
$R$ site but also to form solid solutions \cite{cah19} and fabrication of
high-entropy ceramics \cite{qin21}. Along this line of thinking one may ask
whether the two types of electronic inhomogeneities discussed above could be
combined in a substitution series \seb, and possibly be tailored by adjusting
$x$. Here it should be noted that substitution of Sm by Eu increases the
lattice constant only slightly, but changes the carrier density due to the
different valencies and simultaneously influences the magnetic order
\cite{yeo09}. As a result, an early study of the substitution series focused
on the global magnetic and transport properties \cite{yeo12}. At low
temperature, insulating behavior was found up to $x \lesssim$ 0.4 and metallic
behavior for larger $x$. Antiferromagnetism prevailed for 0.2 $\lesssim x
\lesssim$ 0.95. In contrast, the regions for smaller and larger $x$ resemble
the respective end member: SmB$_6$ is nonmagnetic \cite{men69,ghe19} in its
ground state (it orders magnetically under pressure above 6~GPa \cite{bar05})
like many other Kondo insulators, whereas EuB$_6$ is a ferromagnetic semimetal
\cite{aro99}. A photoelectron spectroscopy study revealed that the Kondo
coherence and the related gap survive at least up to $x = 0.15$ and up to 50~K
\cite{yam13}. Magnetotransport studies on samples with $x \leq$ 0.05 supported
the persistence of a gap \cite{gab16} and the surface states \cite{ani24}. More
recent studies were able to follow the topological coherence encountered in
SmB$_6$ up to $x = 0.3$ which, again, is related to the presence of a direct
Kondo gap \cite{mia21,xu21}.

In contrast, we concentrate on the Eu-rich samples and investigate the changes
of magnetic and magnetotransport properties upon increasing Sm content in
EuB$_6$. As the ferromagnetic Curie temperature $T_{\rm C}$ is expected to
decrease rapidly with $x$ departing from 1 \cite{yeo12}, samples with closely
spaced $x$ were prepared. We find that the fate of the polaron formation in
\seb\ as the Eu content is reduced to below $x = 1$ does not follow this rapid
dependence of $T_{\rm C}(x)$. In consequence, and in line with recent
observations in few other antiferromagnetic Eu compounds \cite{ros20,ale23,
kre23}, we find indications for ferromagnetic (fm) polaron formation in the
globally antiferromagnetic (afm) material Sm$_{0.05}$Eu$_{0.95}$B$_6$. These
results highlight the complexity of the interactions in Eu-based materials.

\section{Materials and Methods}
Single crystalline samples \seb\ investigated here were synthesized by the
Al-flux technique \cite{ros18}. The nominal stoichiometries were chosen to
allow a closer investigation of the Eu-rich samples: $x =$ 0, 0.2, 0.9, 0.95,
0.99 and 1.0.

Magnetic measurements up to applied fields of 7~T were conducted using a
magnetic property measurement systems (MPMS3, Quantum Design Inc.). If not
noted otherwise, the magnetic ac susceptibility $\chi_{\rm ac}$ was measured
with an applied ac field of 10 Oe after cooling in zero applied field (ZFC).
Electronic transport was measured with a standard 4-probe configuration and an
ac technique utilizing a physical property measurement system (PPMS, Quantum
Design Inc.). In some cases, an external, lock-in-based circuitry was hooked
up to the PPMS in an effort to improve accuracy.

\section{Results}
\subsection{Magnetic measurements} \label{sub:mag}
\begin{figure}[t]
\centering
\includegraphics[width=8.8cm]{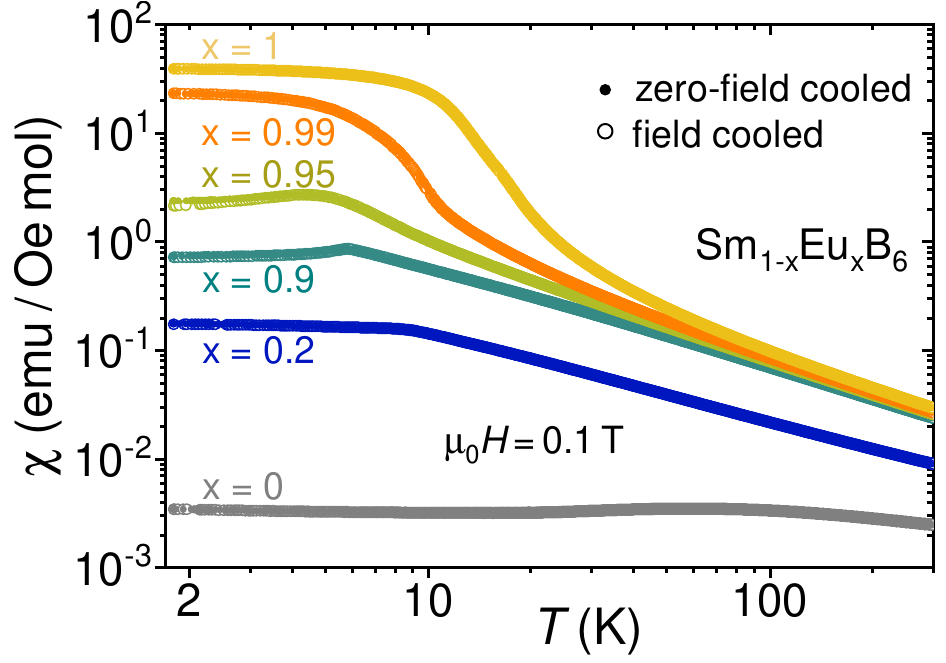}
\caption{Zero-field cooled (filled symbols) and field cooled (open circles)
susceptibility $\chi (T)$ of \seb\ measured in a magnetic field of
$\mu_0 H =$ 0.1~T.} \label{sus-all}
\end{figure}

The dc susceptibilities $\chi (T)$ measured for the differently substituted
samples \seb\ at 0.1~T after zero-field and field cooled conditions are
presented in Fig.\ \ref{sus-all}. The data of pure SmB$_6$ can very nicely be
compared to earlier results \cite{nic71,gab02}. The magnetic properties of Eu
upon substitution of Sm become immediately obvious, even for $x$ as small as
0.2: There is an increase of the low-$T$ susceptibility of almost two orders of
magnitude \cite{yeo12}. In comparison, for a similar amount of La-substitution
$\chi (T)$ does not even increase by a factor of 1.5 \cite{gab02}. The onset of
afm order in sample $x =$ 0.2 is observed at around 8.6~K, in line with
\cite{yeo12}. The latter is reduced to 5.8~K for $x =$ 0.9 and 4.3~K for $x =$
0.95. We note that magnetization measurements for samples $x =$ 0.9 and 0.95
did not reveal signatures of a fm component at zero field (besides the increase
in the magnitude of $\chi$) as will be discussed below.

\begin{figure*}[t]
\centering
\includegraphics[width=12.8cm]{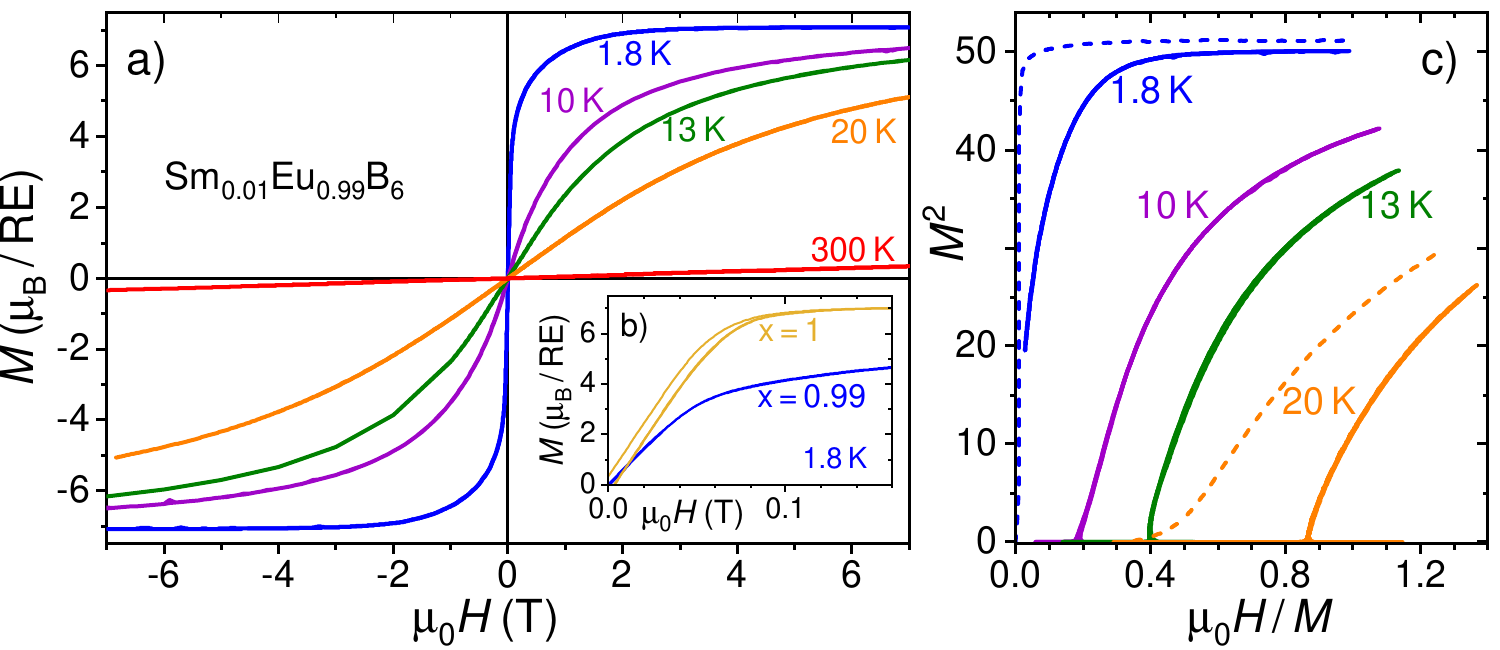}
\caption{(a) Magnetization curves $M(H)$ of Sm$_{0.01}$Eu$_{0.99}$B$_6$ for
different temperatures. (b) Low-field range of the $M(H)$ data at 1.8~K for
Sm$_{0.01}$Eu$_{0.99}$B$_6$ and EuB$_6$. (c) Arrott plot, $M^2$ vs.\ $\mu_0
H/M$, of the $M(H)$ data shown in (a). Dashed lines are data of pristine
EuB$_6$ at 1.8~K and 20~K for comparison.}
\label{mag099}  \end{figure*}

For $x =$ 0.99 and 1.0, a qualitatively different behavior is observed: The
rising $\chi (T)$ down to lowest $T$ indicates ferromagnetic behavior.
The Curie temperatures $T_{\rm C}$ of ferromagnets can be estimated from
Arrott plots \cite{arr57}, as shown in Fig.\ \ref{mag099}(c) for sample $x =
0.99$. In particular, the temperature dependence of the zero-field
susceptibility $\chi_0(T)$ was evaluated, where $\chi_0^{-1}(T)$-values are
the intercepts with the $H/M$-axis \cite{kou64,sue00}. We estimated $T_{\rm C}
\sim$ 7.3~K for $x = 0.99$ and $T_{\rm C} \sim$ 12.7~K for $x = 1.0$. The
latter value of the end member EuB$_6$ is in good agreement with reported ones \cite{sue98,sue00,urb04} while the exemplary $M^2$ vs.\ $\mu_0 H/M$ data,
dashed lines in Fig.\ \ref{mag099}(c), are in accordance with \cite{sue00}.
It is important to note, however, that $T_{\rm C}$ in EuB$_6$ is
different from the metal-semiconductor transition temperature $T_{\rm M}$ at
which the magnetic polarons percolate \cite{nyh97,yu05,zha09,das12}. The
formation of magnetic polarons in EuB$_6$ is reported to set in between 25 --
35~K \cite{nyh97,urb04,poh18,min21}.

Magnetization curves $M(H)$ of Sm$_{0.01}$Eu$_{0.99}$B$_6$ were measured at
several temperatures; the results are presented in Fig.\ \ref{mag099}(a).
The saturation value of $M_{\rm sat} \approx 7.08\; \mu_{\rm B}/$Eu agrees well
with the expected saturation magnetic moment of primarily Eu$^{2+}$, $g
\mu_{\rm B} J =$ 7~$\mu_{\rm B}$. Even at small fields there is no hysteresis
seen at $T =$ 1.8~K, in contrast to the data for EuB$_6$ ($x = 1$, $M_{\rm
sat} \approx 7.02\; \mu_{\rm B}/$Eu), Fig.\ \ref{mag099}(b). We note that care
has to be taken if $M(H)$-data are to be inspected on a small field scale: The
\begin{figure}[b]
\centering
\includegraphics[width=8.6cm]{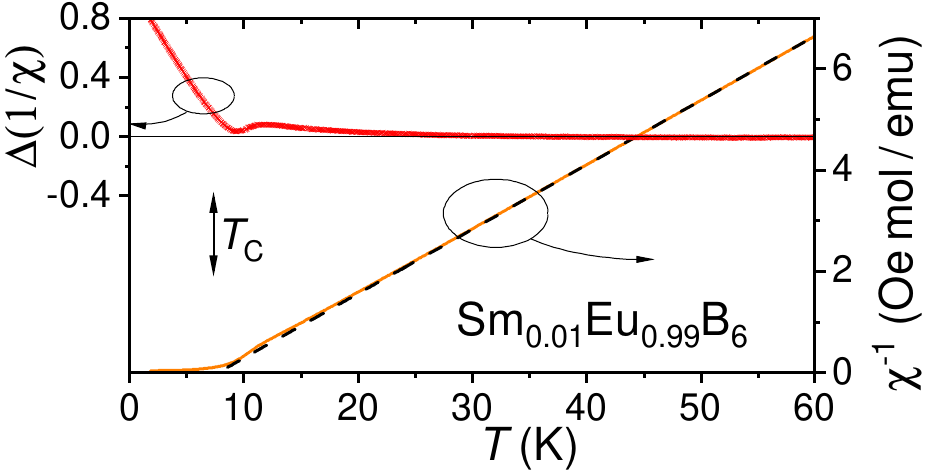}
\caption{Inverse susceptibility $1/\chi(T)$ of Sm$_{0.01}$Eu$_{0.99}$B$_6$
(orange line and right scale) and Curie-Weiss fit (dashed line). $\Delta (1/
\chi)$ (red markers and left scale) denotes the difference between the
$1/\chi$-data and the linear fit (same units as $\chi^{-1}$).}  \label{sus099}
\end{figure}
remnant field of the superconducting magnet of the MPMS was determined before
and after each measurement.

The inverse susceptibility $1/\chi(T)$ of Sm$_{0.01}$Eu$_{0.99}$B$_6$ measured
at a field of 0.1~T is presented in Fig.\ \ref{sus099}. A Curie-Weiss fit of
the data in the range 25~K $\leq T \leq$ 250~K yields a Curie-Weiss
temperature $\theta_{\rm CW} = +7.7 \pm 0.4$~K and an effective moment $\mu
_{\rm eff} = 7.93 \pm 0.08 \,\mu_{\rm B}$, the latter in excellent agreement
with $\mu_{\rm eff}^{\rm cal} = 7.94\,\mu_{\rm B}$ expected for Eu$^{2+}$.
However, there are clear deviations from the Curie-Weiss law below about 20~K
as seen by the difference $\Delta (1/\chi)$ between the $1/\chi$-data and the
extrapolated linear fit, red markers in Fig.\ \ref{sus099}. We speculate that
these deviations result from the impact of magnetic inhomogeneities like
polarons, as observed in a number of Eu compounds \cite{sue00,bea22,mit22,
ale23}.

\begin{figure}[b]
\centering
\includegraphics[width=8.7cm]{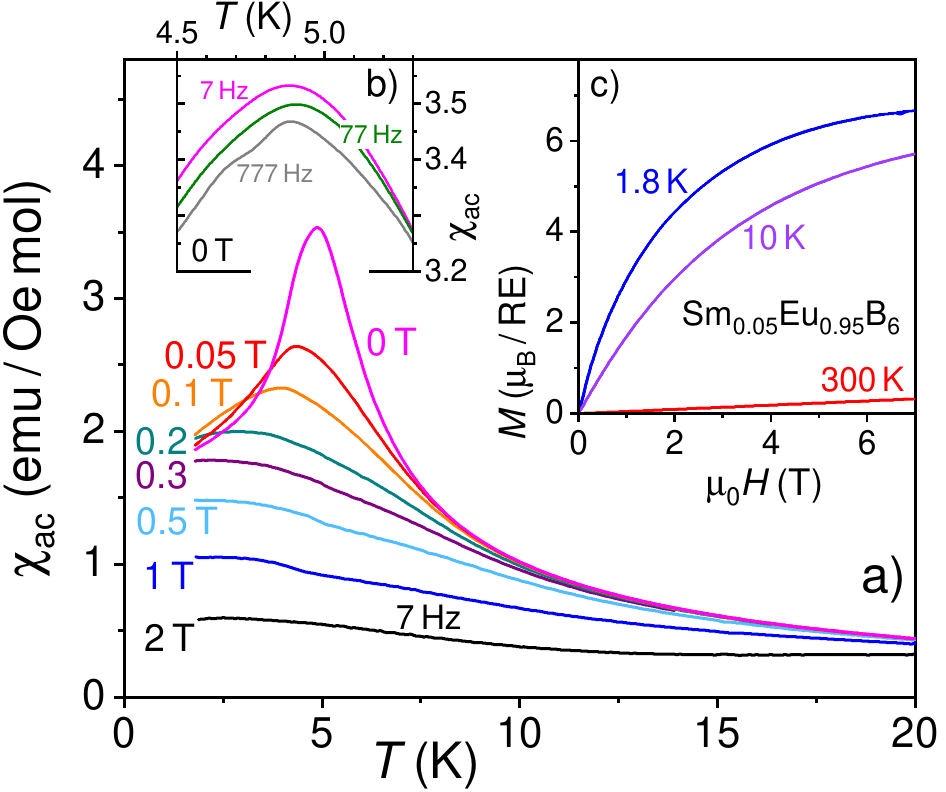}
\caption{(a) Susceptibility $\chi_{\rm ac}(T)$ of Sm$_{0.05}$Eu$_{0.95}$B$_6$
measured at different fields as indicated. (b) Frequency dependence of
$\chi_{\rm ac}(T)$ for zero applied field. (c) Magnetization curves $M(H)$ of
Sm$_{0.05}$Eu$_{0.95}$B$_6$ at selected temperatures.} \label{mag095}
\end{figure}

\begin{figure}[t]
\centering
\includegraphics[width=8.7cm]{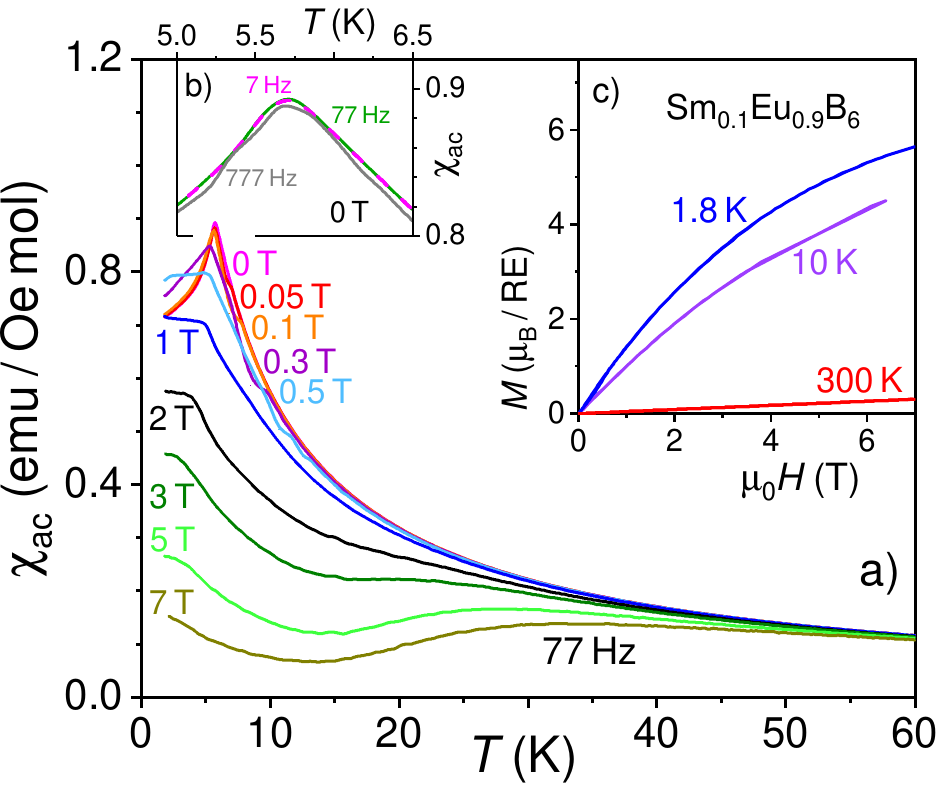}
\caption{(a) $\chi_{\rm ac}(T)$ of Sm$_{0.1}$Eu$_{0.9}$B$_6$ measured at
different magnetic fields. (b) Frequency dependence of $\chi_{\rm ac}(T)$ for
zero applied field. (c) Magnetization curves $M(H)$ of sample
Sm$_{0.1}$Eu$_{0.9}$B$_6$ at 1.8~K, 10~K and 300~K.} \label{mag09}
\end{figure}

We now turn to sample Sm$_{0.05}$Eu$_{0.95}$B$_6$. Figure \ref{mag095}(a)
presents the low-temperature ac susceptibility $\chi_{\rm ac}(T)$ as measured
at different applied magnetic fields. Clearly, an afm transition is indicated
by a peak in $\chi_{\rm ac}(T)$ at $T_{\rm N} = 4.9 \pm 0.1$~K. As expected,
this peak is gradually suppressed by applying a magnetic field. At $\mu_0 H =$
0.1~T, the peak temperature $T_{\rm max} \approx$ 4.1~K of $\chi_{\rm ac}(T)$
is in good agreement with the result shown in Fig.\ \ref{sus-all}. However,
the peak in $\chi_{\rm ac}(T)$ in zero field is not as sharp as in typical
antiferromagnets, see e.g.\ \cite{mug22}. In order to nonetheless exclude the
possibility of a spin-glass-like behavior as, e.g., observed in
Cu$_{1-x}$Mn$_x$ for $0.01 \leq x \leq 0.063$ \cite{mul81}, we measured the
frequency dependence of $\chi_{\rm ac}(T)$ near $T_{\rm N}$, as plotted in
Fig.\ \ref{mag095}(b). A shift of $T_{\rm max}$ with frequency is not obvious
albeit a reduction of $\chi_{\rm ac}(T_{\rm max})$ with increasing frequency
is observed. These observations can be viewed as indications for the existence
of magnetic polarons [in contrast to Sm$_{0.1}$Eu$_{0.9}$B$_6$ in Fig.\
\ref{mag09}(b)]. Additionally, there is a strong reduction of
$\chi_{\rm ac}(T)$ with increasing $H$ well above $T_{\rm N}$, Fig\
\ref{mag095}(a). A similar behavior in Eu$_5$In$_2$Sb$_6$ was also discussed
\cite{ale23} as an indication for polaron formation.

Ferromagnetic interactions in the sample with $x= 0.95$ are corroborated by
\begin{figure}[t]
\centering
\includegraphics[width=7.4cm]{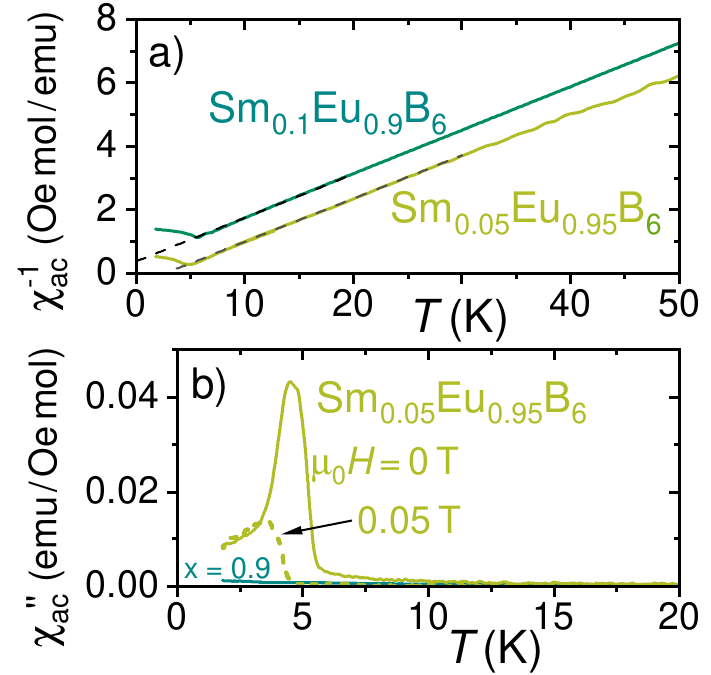}
\caption{(a) Inverse susceptibilities $1 / \chi_{\rm ac}(T)$ of
Sm$_{1-x}$Eu$_x$B$_6$ for $x =$ 0.9 and 0.95. The black dashed lines are
extrapolations of linear fits to the high-temperature data. (b) Imaginary
component $\chi_{\rm ac}''(T)$ for sample $x =$ 0.95 at zero field (line) and
50~mT (dashed). For comparison, $\chi_{\rm ac}''(T)$ of sample $x =$ 0.9 at
zero field is included.} \label{inv-sus}
\end{figure}
further analysing $\chi_{\rm ac}(T)$. Well above $T_{\rm N}$, $\chi_{\rm
ac}(T)$ follows nicely a Curie-Weiss law with a clearly \emph{positive}
$\theta_{\rm CW} = +2.7 \pm 0.5$~K, see Fig.\ \ref{inv-sus}(a). The value
$\mu_{\rm eff} = 7.7 \pm 0.1\, \mu_{\rm B}$ obtained from the fit is in good
agreement with $\mu_{\rm eff}^{\rm cal} = 7.59\,\mu_{\rm B}$ expected for a
mixture of 95\% Eu$^{2+}$ and 5\% Sm$^{3+}$. A large imaginary component of the
susceptibility, $\chi_{\rm ac}''(T)$, is observed near $T_{\rm N}$, Fig.\
\ref{inv-sus}(b), in line with an magnetically inhomogeneous state. An applied
field as small as 50~mT suppresses $\chi_{\rm ac}''(T)$ significantly
indicating an effective suppression of these inhomogeneities. Importantly,
$\chi_{\rm ac}''(T)$ is already enhanced at temperatures well above $T_{\rm N}$
again indicating magnetic inhomogeneities. For comparison, the small $\chi_{\rm
ac}''(T)$ for sample $x =$ 0.9 is also shown in Fig.\ \ref{inv-sus}(b).

Here we recall that sample $x = 0.99$ is a ferromagnet while samples with $x =
0.90$ and 0.95 order globally antiferromagnetically. However, the positive
value of $\theta_{\rm CW}$ and the behavior of $\chi_{\rm ac}''(T)$ of
Sm$_{0.05}$Eu$_{0.95}$B$_6$ may again be indicative of polaron formation, as
will be discussed below. In contrast, data of the afm sample
Sm$_{0.1}$Eu$_{0.9}$B$_6$ do not reveal any sign of polaron formation in
$\chi_{\rm ac}''(T)$, see in Fig.\ \ref{inv-sus}(b). Fitting $\chi_{\rm ac}(T)$
to a Curie-Weiss law yields $\theta_{\rm CW} = -2.9 \pm 0.5$~K and $\mu_{\rm
eff} = 7.6 \pm 0.1\, \mu_{\rm B}$. As shown in Fig.\ \ref{mag09}(a), $T_{\rm N}
= 5.7 \pm 0.1$ K. The peak in $\chi_{\rm ac}(T)$ is also suppressed by magnetic
fields but much less effectively compared to the sample $x = 0.95$. Moreover,
there is no noticeable frequency dependence of $\chi_{\rm ac}(T)$ in Fig.\
\ref{mag09}(b) pointing to a homogeneous magnetic state at zero field (in
contrast to the observations around 3~T discussed below).

\subsection{Magnetotransport measurements} \label{sub:resi}
\begin{figure}[t]
\centering
\includegraphics[width=8.5cm]{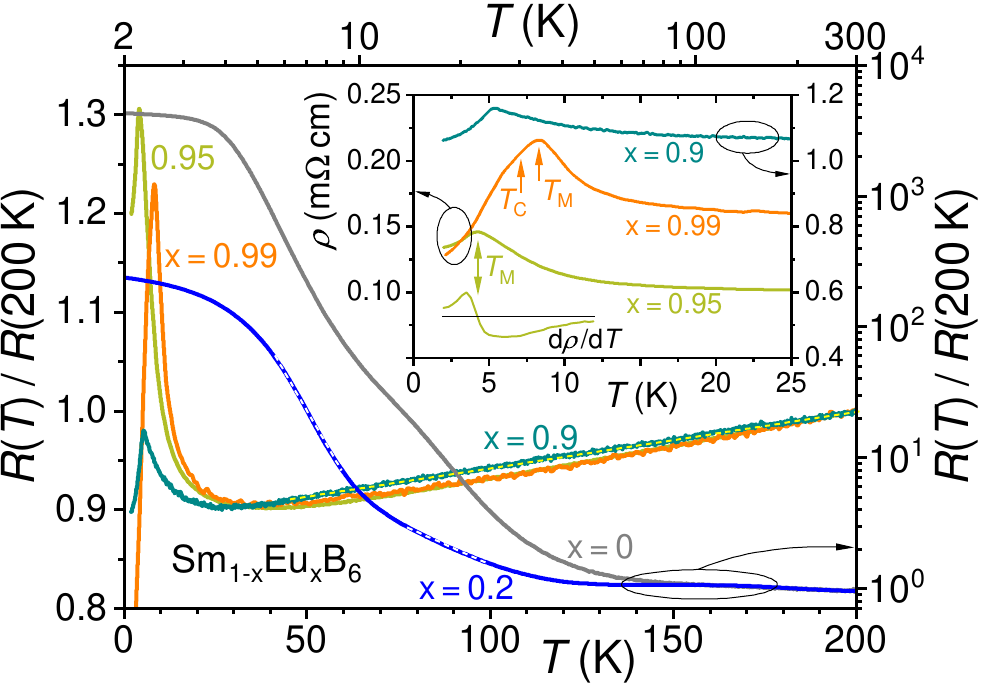}
\caption{Resistance ratio scaled to the respective value at $T =$ 200~K for
samples $x =$ 0, 0.2 (upper and right logarithmic scale) and $x =$ 0.9, 0.95,
0.99 (lower and left linear scale). Dashed lines mark linear behavior. Inset:
Low-temperature resistivities $\rho(T)$ of samples $x=$ 0.95, 0.99 (left scale)
and $x =$ 0.9 (right scale, same unit as left scale). For $x =$ 0.99,
$T_{\rm C}$ from $\chi_{\rm ac}$ and the maximum of $\rho(T)$ at $T_{\rm M}$
are marked. For $x =$ 0.95, the derivative d$\rho / {\rm d}T$ and $T_{\rm M}$
are shown.}  \label{res-all}
\end{figure}

Figure \ref{res-all} summarizes the temperature dependent resistance behavior
of our samples. The resistance of Sm-rich samples $x =$ 0 and 0.2 increases
upon lowering $T$ indicating insulating behavior. For pristine SmB$_6$, two
temperature ranges with different values of the hybridization gap $\Delta$ can
be distinguished in resistance $R(T) \propto \exp(\Delta / k_{\rm B}T)$
\cite{eo19,ale21b} where $k_{\rm B}$ is the Boltzmann constant. The values
obtained here, $\Delta_1 \sim$ 2.6~meV at lower $T$ and $\Delta_2 \sim$ 5.2~meV
at higher $T$, are in good agreement with the results in \cite{ale21b}. For
$x =$ 0.2 there are also two distinct temperature ranges (white dashed lines in
Fig.\ \ref{res-all}) albeit of narrower extent in $T$. Interestingly, upon Eu
substitution with $x =$ 0.2 the low-$T$ gap $\Delta_1 \sim$ 2.7~meV remains
stable, while the high-$T$ gap $\Delta_2 \sim$ 1.9~meV reduces drastically. In
addition, $R(T)$ appears to approach saturation at low $T$, in line with the
reported existence of surface states at this substitution level \cite{mia21,
xu21} but in contrast to similar La, Lu or Ce substitution \cite{gab16,yam13,
hat20}.

Higher Eu substitutions $x =$ 0.9, 0.95 and 0.99 result in semimetallic
behavior (as in EuB$_6$ \cite{kun04}). In fact, $\rho(T)$ of sample $x =$ 0.99
is highly reminiscent of pristine EuB$_6$ \cite{sue98,das12}: it passes through
a minimum at $T^* \sim$ 35~K before steeply rising to a peak at $T_{\rm M}
\approx$ 8.4~K. $T_{\rm C}$ (as determined from $\chi_{\rm ac}(T)$) manifests
itself as a small hump in $\rho (T)$, as seen in the inset of Fig.\
\ref{res-all}. These close similarities to EuB$_6$ \cite{sue00,das12}
encourage us to assign $T^*$ to the polaron formation and $T_{\rm M}$ to their
percolation. We note, however, $T_{\rm M}$ and $T_{\rm C}$ in sample $x =$ 0.99
are reduced to almost half the values found in EuB$_6$.

In general, sample $x =$ 0.95 exhibits a $\rho(T)$-behavior very similar to
sample $x =$ 0.99, specifically if plotted as $R(T)/ R(T\! =\,$200~K),
irrespective of the different magnetic order at low $T$ in these two samples.
\begin{table}[b]
\caption{Properties of the highly Eu-containing samples Sm$_{1-x}$Eu$_x$B$_6$.}
\begin{ruledtabular}
\begin{tabular}{c|cccccc}
Sample & order & $T_{\rm C}$/$T_{\rm N}$ & $\theta_{\rm CW}$ & $\mu_{\rm eff}$
& $M_{\rm sat}$ & $T_{\rm M}$ \\
$x$  &         & K   & K       & $\mu_{\rm B}$ & $\mu_{\rm B}$ & K \\ \hline
0.9  & afm     & 5.7 & $-2.9$  & 7.6  &  /   & 5.4 \\
0.95 & afm     & 4.9 & $+2.7$  & 7.7  & 6.91 & 4.2 \\
0.99 &  fm     & 7.3 & $+7.7$  & 7.93 & 7.08 & 8.4 \\
1.0  &  fm     &12.7 & $+15.4$ & 8.1  & 7.02 & $15.2\,$\cite{das12} \\ \hline
error& & $\pm 0.1$ & $\pm 0.5$ & $\pm0.1$   & $\pm 0.1$ & $\pm 0.2$
\end{tabular} \end{ruledtabular}
\label{tab-prop}
\end{table}
We recall that $T_{\rm N} =$ 4.9~K for sample $x =$ 0.95. As seen from the
inset to Fig.\ \ref{res-all}, $T_{\rm M} \approx$ 4.2~K is now below $T_{\rm
N}$ and the only subtle hint at the onset of afm order in $\rho(T)$ might be a
broadened minimum in d$\rho(T)/$d$T$. A subtle influence of the afm order on
$\rho(T)$ is suggested by the $\rho(T)$-behavior observed for sample $x =$ 0.9:
A peak of much less pronounced relative height in $R(T)/ R(T\! =\,$200~K) is
observed at $T_{\rm M} \approx$ 5.4~K, i.e.\ very close to $T_{\rm N} =$ 5.7~K,
and hence, likely related to the onset of afm order.

It is interesting to note that $\rho(T)$ of sample Sm$_{0.1}$Eu$_{0.9}$B$_6$
follows very nicely a linear dependence within the range 45~K $\lesssim T
\lesssim$ 220~K, see yellow dashed line in Fig.\ \ref{res-all}. EuB$_6$ has a
low carrier concentration of order 0.01 per unit cell in the fm and
paramagnetic regime \cite{aro99,zha08} and hence, a scattering model as
proposed in \cite{hwa19} may be relevant assuming an only moderately enhanced
carrier concentration due to substitution in Sm$_{0.1}$Eu$_{0.9}$B$_6$.
However, additional measurements on the latter material are needed, e.g.\ of
the Hall effect or the impact of disorder \cite{pat23}, to gain further insight
on the linear-in-$T$ resistivity.

One of the striking phenomena related to polaron physics is an enhanced
negative magnetoresistance (MR) \cite{ter97,coe99,mol07}. For separated
polarons in EuB$_6$, the charge transport is dominated by magnetic scattering
at these fm polarons \cite{glu07}. Above $T_{\rm M}$, the more conducting
volumes of the fm polarons grow out of the less conductive para\-magnetic
``background'' either upon lowering the temperature or by application of a
magnetic field \cite{sue00,zha08,zha09,das12}. Upon reaching a high enough
number density (in zero field this is at $T_{\rm M}$) the polarons percolate
and the resistivity drops giving rise to a large negative MR. Obviously, at
higher magnet fields the percolation threshold becomes shifted to higher
temperatures.

\begin{figure*}[t]
\centering
\includegraphics[width=12.2cm]{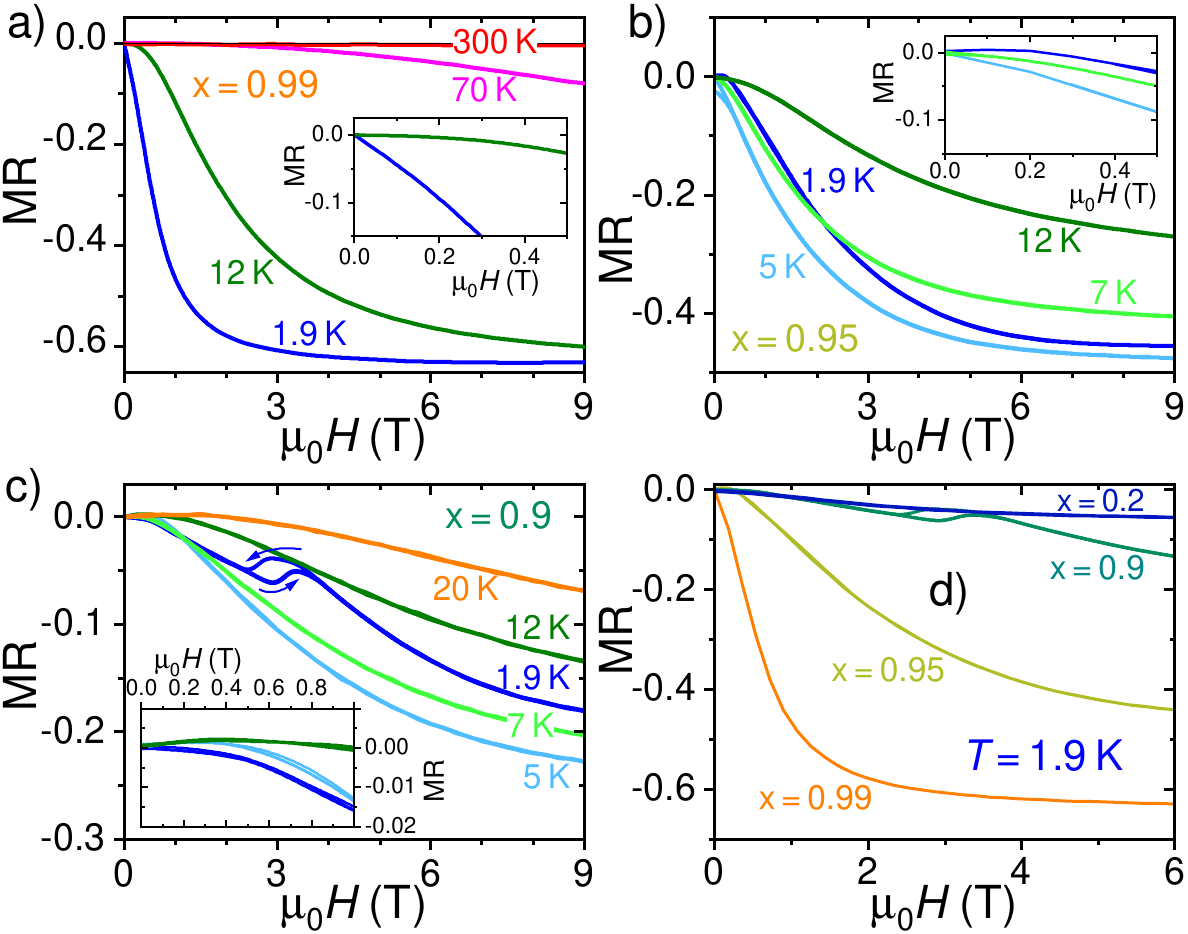}
\caption{Magnetoresistance MR for samples (a) $x=$ 0.99, (b) $x=$ 0.95 and (c)
$x=$ 0.9 at selected temperatures. Insets: Zoom into the low-field region (same
color codes as main panels). (d) Comparison of MR at $T =$ 1.9~K for the
differently substituted samples.} \label{MR-all}
\end{figure*}

\begin{figure}[b]
\centering
\includegraphics[width=6.8cm]{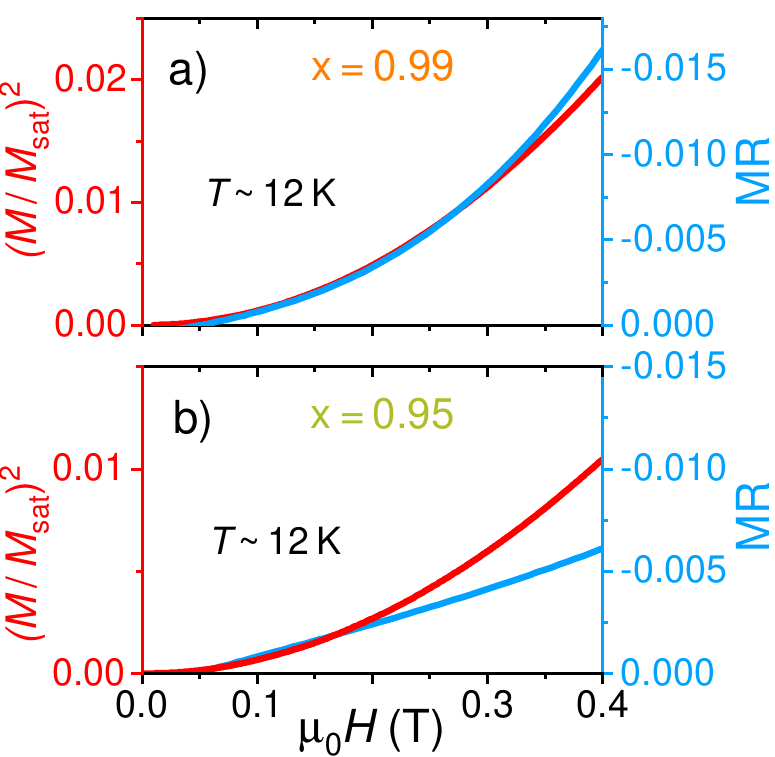}
\caption{Squared relative magnetization $(M(H) / M_{\rm sat})^2$ (red) and MR
(blue) for samples (a) $x=$ 0.99 and (b) $x=$ 0.95 at small fields and in the
polaronic regime ($\sim$12~K).}
\label{M2MR}  \end{figure}

At first glance, $\rho(T)$ of our sample $x=$ 0.99 is qualitatively very
similar to pristine EuB$_6$ \cite{sue00}, with $T_{\rm C}$, $T_{\rm M}$ (see
Table \ref{tab-prop}) and $\rho(T_{\rm M})$ somewhat reduced. The latter is
likely the result of an increased carrier density due to the substitution of Eu
by Sm. The MR, however, is notably different: While MR $= \left[\rho(H) -
\rho(H\! = \! 0)\right] / \rho(H\! = \! 0)$ is positive for EuB$_6$ below about
5~K \cite{aro99,pas00}, i.e.\ well in the metallic regime, this is \emph{not}
observed for sample $x=$ 0.99, see Fig.\ \ref{MR-all}(a). The positive MR in
the fm regime of EuB$_6$ has been explained by the subtle interplay of electron
and hole carriers in a two-band model whereas the negative MR above $T_{\rm M}$
was attributed to spin disorder scattering \cite{cal04,bat06,zen22} near
$T_{\rm C}$ and percolation of magnetic polarons at intermediate temperatures
further above $T_{\rm C}$ \cite{zha09,poh18,das12}, both known to result in
large negative magnetoresistance \cite{maj98}. Therefore, we interpret the large
negative MR even at lowest $T =$ 1.9~K (MR $= -0.47$ at 1~T) also in terms of
predominant suppression of spin disorder scattering in our 1\%~Sm-substituted
sample. Strikingly, very similar values at which the MRs saturate are observed
for high fields: at 1.9~K we find MR~$= -0.63$, compared to an estimate of
$-0.62$ at $T =$ 12~K, the latter being well above $T_{\rm C}$ and $T_{\rm M}$,
i.e.\ in the polaron regime. Moreover, for spin disorder scattering a scaling
was suggested \cite{fur94,maj98b} for small magnetization: MR $= -C (M / M_{\rm
sat})^2$. As shown in Fig.\ \ref{M2MR}(a), this scaling is obeyed in our sample
$x =$ 0.99 up to $(M / M_{\rm sat})^2 \lesssim$ 0.012 with $C \sim 0.7$. The
latter is much smaller than the value reported for EuB$_6$ ($C =$ 75,
\cite{sue00}) yet close to the prediction 1 $\leq C \lesssim$ 5 \cite{fur94}.
We also note that electrons are thought to be responsible for the magnetic
ordering via RKKY interaction \cite{cal04}. A changing carrier density will
affect the RKKY interaction strength. Hence, one may speculate that 
substitution of Eu by 1\% Sm helps to tip the subtle balance between hole and 
electron carriers in EuB$_6$ toward electrons in Sm$_{0.01}$Eu$_{0.99}$B$_6$.

The afm sample $x =$ 0.95 exhibits also large negative MR values reaching
almost $-0.5$. Spin disorder scattering in the paramagnetic regime is again
indicated by the scaling of MR with $M^2$ in Fig.\ \ref{M2MR}(b). For this
sample, the scaling works up to only $(M / M_{\rm sat})^2 \lesssim$ 0.003 which
is likely related to the lesser increase of $M(H)$ with $H$ compared to sample
$x=$ 0.99 (see, e.g., the magnetization data at 10~K in Figs.\ \ref{mag095}(c)
and \ref{mag099}(a), respectively). Accordingly, $C$ is increased to about
0.94. This points to magnetic inhomogeneities being also present in sample $x=$
0.95. More striking, however, is the evolution of MR$(H)$ with temperature in
Fig.\ \ref{MR-all}(b): $|$MR$(H)|$ is larger at all fields for $T =$ 5~K
compared to the data at 1.9~K. Very likely, this is a consequence of the
5~K-curve taken above but very close to $T_{\rm N}$. It is a hallmark of
polaronic systems to exhibit the largest MR values close to the magnetic
ordering temperature \cite{coe99}; together with the large MR values we
therefore surmise that polarons also form in sample $x =$ 0.95. At 1.9~K, the
formation of fm polarons competes with the global afm order and, in result, a
tiny positive MR is observed at small fields, inset to Fig.\ \ref{MR-all}(b),
as also seen in other afm materials with polaron formation \cite{sho19}.

Compared to sample $x =$ 0.95, the MR in sample $x =$ 0.9 is reduced by a
factor of approximately 2, with the exception of the 1.9~K-curve. Intriguingly,
the latter also shows a significant hysteresis near 3~T, Fig.\ \ref{MR-all}(c),
which may point to fm contributions in this afm material. Here it should be
noted that the MR was measured for fields sweeps $\mu_0 H =$ 0~T $\rightarrow
+7$~T $\rightarrow -7$~T $\rightarrow +7$~T. The lack of any difference in the
up-sweep data (curved blue arrows) indicates that this hysteresis is
reproducible and independent of magnetic history. In Ref.\ \cite{daw24}, a
phenomenological model was introduced which predicts a stabilization of fm
polarons inside a globally antiferromagnetically ordered phase upon lowering
the temperature. As one may expect, such polarons are also energetically
favored by the application of a magnetic field. One may therefore speculate
that the hysteresis in MR results from magnetic inhomogeneities that may form
at intermediate fields between the global afm order at low fields and the
field-polarized state at high fields. The small slope of MR$(H)$ at low fields
[similar in magnitude to the one observed for afm sample $x=$ 0.2 up to
$\sim$1.3~T, Fig.\ \ref{MR-all}(d)], which increases drastically above the
hysteresis, may support such a line of thinking.

\begin{figure}[t]
\centering
\includegraphics[width=8.6cm]{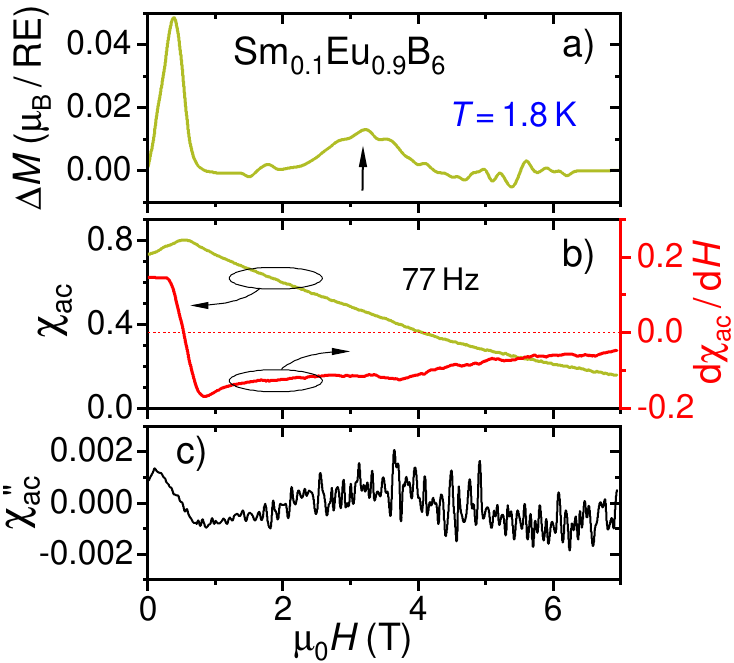}
\caption{Detailed magnetic measurements at $T =$ 1.8~K for sample
Sm$_{0.1}$Eu$_{0.9}$B$_6$. (a) $\Delta M = M(H\!\!\downarrow) - M(H\!\!
\uparrow)$, i.e.\ the difference between down- and up-sweep magnetization data
of Fig.\ \ref{mag09}(c). The arrow marks the maximum of the high-field
hysteresis in magnetization. (b) Field dependence of $\chi_{\rm ac}(H)$ (left
scale) and its derivative (right scale), cf.\ Fig.\ \ref{mag09}(a). (c)
Imaginary component $\chi_{\rm ac}^{''}$. All susceptibilities are in units of
(emu/Oe$\,$mol).} \label{Eu09magn}
\end{figure}

In order to scrutinize the possible existence of magnetic polarons in
Sm$_{0.1}$Eu$_{0.9}$B$_6$ at intermediate fields, the results of magnetic
measurements were analyzed in detail. Figure \ref{Eu09magn}(a) exhibits the
difference \mbox{$\Delta M = M(H\!\!\downarrow)$} $- M(H\!\!\uparrow)$ of the
magnetization measured while sweeping the magnetic field up $(H\!\!\uparrow)$
and down $(H\!\!\downarrow)$. Clearly, there is some hysteresis in $M(H)$ [not
seen at the scale of Fig.\ \ref{mag09}(c)] centered around 3~T, as in the
MR$(H)$, see arrow in Fig.\ \ref{Eu09magn}(a). Apparently, the magnetization
within the volume of these magnetic inhomogeneities increases the total
magnetization only slightly despite introducing significant scattering. Both,
$\chi_{\rm ac}(H)$ and $\chi_{\rm ac}^{''}(H)$ also exhibit tiny humps just
above 3~T, Figs.\ \ref{Eu09magn}(b) and (c). Here we wish to point out that
the accuracy of our measurement setup is, unfortunately, of the order
0.005~$\mu_{\rm B}$/RE which may explain the noise of our $\chi_{\rm
ac}^{''}$-data as well as their seemingly negative values at certain magnetic
fields. Notably, there appears another feature in the magnetic data at small
fields (below 0.6~T) which is not seen in the MR [cf.\ inset to Fig.\
\ref{MR-all}(c)] and whose origin is presently unknown.

\section{Discussion}
The fm order in EuB$_6$ is very efficiently suppressed by introducing Sm: only
1\% Sm reduces $T_{\rm C}$ to almost half its value, see Table \ref{tab-prop}.
For comparison, Eu$_{0.8}$Ca$_{0.2}$B$_6$ still orders ferromagnetically albeit
with reduced $T_{\rm C}$ \cite{pas00,rhy03}. In contrast, our sample
Sm$_{0.05}$Eu$_{0.95}$B$_6$ orders antiferromagnetically with $T_{\rm N} =$
4.9~K even though the small but positive Curie-Weiss temperature
$\theta_{\rm CW}$ indicates the presence of fm interactions. The latter likely
support the formation of fm polarons in our sample $x =$ 0.95. Such polarons
are indicated by a large negative MR which is enhanced near the magnetic
ordering temperature. More generally, in EuB$_6$ the fm and afm interactions
appear to be close in energy, and to compete. Hence, only small substitutions
of Eu by Sm are then sufficient to change the ground state from ferromagnetic
in EuB$_6$ to antiferromagnetic in $x \leq 0.95$ while polaron formation is
still feasible.

Indeed, indications for the formation of fm polarons are also found in the MR
of sample Sm$_{0.1}$Eu$_{0.9}$B$_6$ but only within a field range of
approximately 2.3 -- 3.6~T. This is consistent with a recently developed
phenomenological model which predicts a stabilization of fm polarons with
increasing magnetic field within a material with global afm order if there is
an appropriately low charge carrier concentration and a sufficiently strong
exchange interaction between the local RE spin and the spin of the charge
carriers \cite{mol07,daw24}.

\vspace*{0.4cm}
\noindent \textbf{Author Contributions:}
Conceptualization, S.W., J.M. and Z.F.; Resources, P.F.S.R.\ and Z.F.;
Investigation, M.V.A.C.; Methodology: S.W.\ and P.S.; Writing---original draft,
S.W.\ and P.S.; Writing---review and editing, S.W., M.V.A.C., P.F.S.R., J.M.\
and P.S. All authors have read and agreed to the published version of the
manuscript.

\vspace*{0.2cm}
\noindent \textbf{Funding:}
Work at the Max-Planck-Institute for Chemical Physics of Solids in Dresden
and at Goethe University Frankfurt was supported by the Deutsche
Forschungsgemeinschaft (DFG, German Research Foundation), Project No.\
449866704. Work at Los Alamos National Laboratory was performed under the
auspices of the U.S.\ Department of Energy, Office of Basic Energy Sciences,
Division of Materials Science and Engineering.

\vspace*{0.2cm}
\noindent \textbf{Acknowledgments:}
S.W. acknowledges insightful discussions with Ulrich Schwarz.

\vspace*{0.2cm}
\noindent \textbf{Data Availability Statement:}
The data presented in this study are available upon reasonable request from
the corresponding author.

\end{document}